\documentclass[useAMS]{mn2e}
 \usepackage{times}
 \usepackage{epsfig}
 \usepackage{amssymb}

 \title[On the origin of the Chelyabinsk superbolide]
       {The Chelyabinsk superbolide: a fragment of asteroid 2011 EO$_{40}$?}

 \author[C. de la Fuente Marcos and R. de la Fuente Marcos]
        {C.~de~la~Fuente~Marcos\thanks{E-mail: nbplanet@fis.ucm.es}
         and
         R. de la Fuente Marcos \\
         Universidad Complutense de Madrid,
         Ciudad Universitaria, E-28040 Madrid, Spain}
 \date{Accepted 2013 July 30.
       Received 2013 July 9;
       in original form 2013 June 11}
 \pubyear{2013}
 
\begin{document}
  \maketitle

  \begin{abstract}
     Bright fireballs or bolides are caused by meteoroids entering the Earth's 
     atmosphere at high speed. Some have a cometary origin, a few may have 
     originated within the Venus-Earth-Mars region as a result of massive 
     impacts in the remote past but a relevant fraction is likely the result of 
     the break-up of asteroids. Disrupted asteroids produce clusters of 
     fragments or asteroid families and meteoroid streams. Linking a bolide to 
     a certain asteroid family may help to understand its origin and pre-impact 
     dynamical evolution. On 2013 February 15, a superbolide was observed in the 
     skies near Chelyabinsk, Russia. Such a meteor could be the result of the 
     decay of an asteroid and here we explore this possibility applying a 
     multistep approach. First, we use available data and Monte Carlo 
     optimization (validated using 2008 TC$_{3}$ as template) to obtain a robust 
     solution for the pre-impact orbit of the Chelyabinsk impactor ($a$ = 1.62 
     au, $e$ = 0.53, $i$ = 3$\fdg$82, $\Omega$ = 326$\fdg$41 and $\omega$ = 
     109$\fdg$44). Then, we use this most probable orbit and numerical analysis 
     to single out candidates for membership in, what we call, the Chelyabinsk 
     asteroid family. Finally, we perform $N$-body simulations to either confirm 
     or reject any dynamical connection between candidates and impactor. We find 
     reliable statistical evidence on the existence of the Chelyabinsk cluster. 
     It appears to include multiple small asteroids and two relatively large 
     members: 2007 BD$_{7}$ and 2011 EO$_{40}$. The most probable parent body 
     for the Chelyabinsk superbolide is 2011 EO$_{40}$. The orbits of these 
     objects are quite perturbed as they experience close encounters not only 
     with the Earth--Moon system but also with Venus, Mars and Ceres. Under such 
     conditions, the cluster cannot be older than about 20--40 kyr.
  \end{abstract}


  \section{Introduction}
     The decay of asteroids in the main belt region is one of the sources of small near-Earth asteroids or meteoroids. The shattered pieces 
     resulting from the collisional, tidal or rotational break-up of a rubble-pile asteroid can spread along the entire orbit of the parent 
     body on a time-scale of hundreds of years (T\'oth, Vere\v{s} \& Korno\v{s} 2011). These meteoroid streams can cause meteor showers on 
     the Earth when their paths intersect that of our home planet (e.g. Jopek \& Williams 2013). Exceptionally bright meteors are popularly 
     known as fireballs. More properly, relatively small impacting objects entering the Earth's atmosphere at high speed and reaching an 
     apparent magnitude of -14 or brighter are called bolides; if the magnitude is -17 or brighter they are known as superbolides (Ceplecha 
     et al. 1999). Superbolides can produce very powerful ballistic shock waves as they move at hypersonic speeds and explosive shock waves 
     when they fragment in the atmosphere or hit the ground to form an impact crater (e.g. Ens et al. 2012). They are also parents of 
     meteorite showers as meteorite-dropping bolides, seeding the ground with fragments of extraterrestrial material (e.g. Foschini 2001). 
     Although not capable of triggering global devastation, they are powerful enough to provoke a significant amount of local damage. These 
     events are not exclusive of the Earth but have also been predicted (Dycus 1969; Adolfsson, Gustafson \& Murray 1996; Christou \& Beurle 
     1999; Bland \& Smith 2000; Christou 2004, 2010; Domokos et al. 2007; Christou, Vaubaillon \& Withers 2008; Christou et al. 2012) and 
     observed (Selsis et al. 2005; Christou, Vaubaillon \& Withers 2007; Hueso et al. 2010; Daubar et al. 2013) in other planets. The 
     connection between asteroidal debris and bolides was first proposed by Halliday (1987), further explored by Williams (2002, 2004) and 
     Jenniskens (2006) and first confirmed observationally by Trigo-Rodr\'{\i}guez et al. (2007). Since then, new examples of bolides 
     associated with asteroids have been reported (Trigo-Rodr\'{\i}guez et al. 2009, 2010; Madiedo et al. 2013).
     
     On 2013 February 15, 03:20:33 GMT a superbolide was observed in the skies near Chelyabinsk, Russia. The event is believed to have been 
     caused by a relatively small impacting object (17--20 m) entering the Earth's atmosphere at high speed and a shallow 
     angle.\footnote{http://neo.jpl.nasa.gov/news/fireball\_130301.html} Calculations by Adamo (2013), Borovicka et al. (Green 2013), 
     Chodas \& Chesley,$^{1}$ Emel'yanenco et al.,\footnote{http://www.inasan.ru/eng/asteroid\_hazard/chelyabinsk\_bolid\_new.html}
     Lyytinen,\footnote{http://www.amsmeteors.org/2013/02/large-daytime-fireball-hits-russia/}  
     Lyytinen \& Matson,\footnote{http://www.projectpluto.com/temp/chelyab.htm} Nakano,\footnote{http://www.icq.eps.harvard.edu/CHELYABINSK.HTML}  
     Proud (2013), Zuluaga \& Ferrin\footnote{http://arxiv.org/abs/1302.5377} and Zuluaga, Ferrin \& Geens 
     (2013)\footnote{http://astronomia.udea.edu.co/chelyabinsk-meteoroid/} revealed that the parent object was one of the Apollo asteroids 
     that periodically cross the orbit of the Earth (see Table \ref{orbits}). In this Letter, we assume that the meteoroid responsible for 
     the Chelyabinsk event was the result of a relatively recent asteroid break-up event and use numerical analysis to single out candidates 
     to be the parent body or bodies. Then we perform $N$-body calculations to further study any possible dynamical connection between the 
     candidates and the superbolide. Our analysis indicates that the Chelyabinsk impactor was a small member of a not-previously-identified 
     young asteroid family. The most probable pre-impact orbit is obtained in Section 2 using Monte Carlo optimization techniques. The 
     candidate selection procedure is described and available information on the candidate bodies is presented in Section 3. The results of 
     our $N$-body calculations together with the numerical model are shown in Section 4. The proposed new asteroid family is characterized 
     in Section 5. Results are discussed and conclusions are summarized in Section 6.
%
%
     \begin{table*}
      \centering
      \fontsize{8}{11pt}\selectfont
      \tabcolsep 0.1truecm
      \caption{Currently available solutions for the pre-impact orbit of the Chelyabinsk superbolide. The solution obtained in this work is 
               also included and it matches that of Nakano. Standard deviations are provided when known. See the text for details.}
      \begin{tabular}{ccccccccc}
       \hline
          Authors                 & $a$ (AU)          & $e$               & $i$ ($^{\circ}$) & $\Omega$ ($^{\circ}$) & $\omega$ ($^{\circ}$) & 
          $P_{\rm 0.050 au}$ & $P_{\rm 0.00004263 au}$ & $\beta_{\rm max}$ \\
       \hline
          Adamo (2013)            & 1.60              & 0.53              & 4.07             & 326.46                & 109.36                &
          0.950\,8           & 0.000\,0633             & 0.986\,2          \\
          Borovicka et al.        & 1.55$\pm$0.07     & 0.50$\pm$0.02     & 3.6$\pm$0.7      & 326.41                & 109.7$\pm$1.8         &
          0.889\,1           & 0.000\,1696             & 0.989\,9          \\
          Chodas \& Chesley$^1$   & 1.73              & 0.57              & 4.2              & -                     & -                     &
          -                  & -                       & -                 \\
          Emel'yanenco et al.$^2$ & 1.77              & 0.58              & 4.3              & -                     & -                     &
          -                  & -                       & -                 \\
          Lyytinen$^3$            & 1.66              & 0.52              & 4.05             & 326.43                & 116.0                 &
          0.336\,4           & 0.000\,0018             & 0.882\,8          \\
          Lyytinen \& Matson$^4$  & 1.660\,079129     & 0.524\,09458      & 4.235\,664       & 326.457\,642          & 114.669\,325          &
          0.330\,2           & 0.000\,0014             & 0.876\,5          \\
          Nakano$^5$              & 1.622\,3665       & 0.531\,1191       & 3.871\,28        & 326.425\,24           & 109.708\,44           &
          0.899\,9           & 0.000\,1183             & 0.985\,8          \\
          Proud (2013)            & 1.47$^{+0.03}_{-0.13}$ & 0.52$^{+0.01}_{-0.05}$ & 4.61$^{+2.58}_{-2.09}$ & 326.53$^{+0.01}_{-0.0}$ & 96.58$^{+2.94}_{-1.72}$ &
          0.768\,1           & 0                       & 0.058\,2          \\
          Zuluaga \& Ferrin$^6$   & 1.73$\pm$0.23     & 0.51$\pm$0.08     & 3.45$\pm$2.02    & 326.70$\pm$0.79       & 120.62$\pm$2.77       &
          0.341\,6           & 0.000\,002              & 0.743\,0          \\
          Zuluaga et al. (2013)   & 1.27$\pm$0.05     & 0.44$\pm$0.02     & 3.0$\pm$0.2      & 326.54$\pm$0.08       & 95.1$\pm$0.8          &
          0.978\,3           & 0                       & 0.451\,3          \\
          Zuluaga et al.$^7$      & 1.368$\pm$0.006   & 0.470$\pm$0.010   & 4.0$\pm$0.3      & 326.479$\pm$0.003     & 99.6$\pm$1.3          &
          1                  & 0                       & 0.203\,0          \\
       \hline
          This work               & 1.623\,75$\pm$0.000\,14 & 0.532\,79$\pm$0.000\,11 & 3.817$\pm$0.005  & 326.409\,0$\pm$0.000\,7   & 109.44$\pm$0.03       &
          1                  & 0.020\,446              & 0.999\,7          \\
       \hline
      \end{tabular}
      \label{orbits}
     \end{table*}
%
%

  \section{Before impact: a Monte Carlo approach}
\begin{figure}
        \centering
        \includegraphics[width=\linewidth]{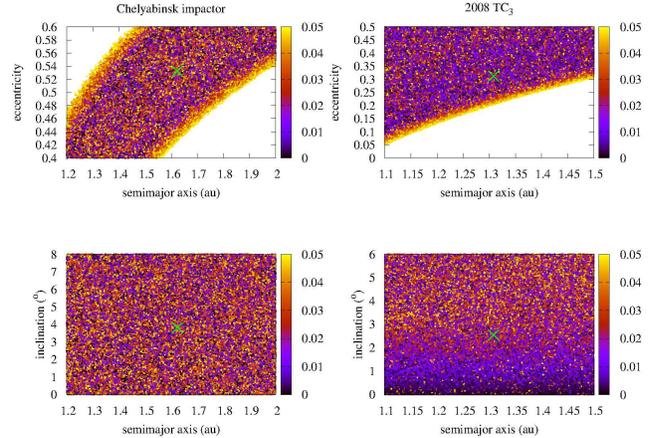}
        \caption{Results from the Monte Carlo approach described in the text for the Chelyabinsk impactor (left-hand panels) and 
                 2008~TC$_{3}$ (right-hand panels). The colours in the colour maps are proportional to the value of the MOID in au for the
                 associated test orbit following the associated colour box. Our preferred solutions appear as green $\times$ signs. The 
                 ranges for the various orbital parameters are indicated in the figures; both the longitudes of the ascending node, $\Omega$, 
                 and the arguments of perihelion, $\omega$, $\in$ (0, 360)$^{o}$.} 
        \label{moid}
     \end{figure}

     If the Chelyabinsk superbolide was the result of the decay of a larger asteroid and we want to identify the putative parent body or 
     bodies, the first step is having a well-defined, statistically robust impactor orbit prior to its collision. Unfortunately, the range 
     of orbital parameters from the solutions provided by the various authors (see Table \ref{orbits}) is too wide to be useful in a 
     systematic search. The impactor came from the direction of the rising Sun and no pre-impact observations have been released yet. The 
     well-known and undeniable facts are that the pre-impact path of the parent body of the Chelyabinsk impactor intersected that of the 
     Earth on 2013 February 15, 03:20:33 GMT and a collision took place. Obviously, for two objects moving in arbitrary Keplerian elliptical 
     orbits to collide, the two paths must intersect and both objects must be in the same spot at the same time. The solution to this 
     problem is not trivial, it does not have an analytical form and is connected with that of finding the minimal distance between two 
     Keplerian orbits (see e.g. Kholshevnikov \& Vassiliev 1999; Gronchi \& Valsecchi 2013). Given two arbitrary orbits, the problem of 
     finding whether or not they intersect and when is, however, well suited for a brute-force Monte Carlo approach in which the two orbits 
     are extensively sampled in phase space and the distance between any two points on the orbits is computed so the minimal distance is 
     eventually found. Let us consider the particular case of an actual collision between a large body with a well-known elliptical orbit 
     and a small object moving in a relatively or completely unknown orbit. If the collision time and the dimensions of the large body are 
     well known, a Monte Carlo calculation can, in principle, help us to determine the trajectory of the small object prior to the collision. 
     It is just a matter of computing the minimal distance between an ensemble of Keplerian test orbits and a set of points compatible with 
     the volume of space occupied by the large body at the collision time. In our implementation, we use the two-body problem expressions in 
     Murray \& Dermott (1999) to generate the orbits with orbital elements obtained, Monte Carlo style, assuming meaningful ranges for the 
     semimajor axis, $a$, the eccentricity, $e$, the inclination, $i$, and both the longitude of the ascending node, $\Omega$, and the 
     argument of perihelion, $\omega$. This method is computer-intensive but if the results can be properly ranked, then an optimal solution 
     for the minimum distance and its associated orbit can be found. Using this Monte Carlo approach, we compute the minimum orbit 
     intersection distance (MOID) for billions of test orbits using a sampling resolution of a few million points per orbit. Neglecting 
     gravitational focusing and in order to have a physical collision, the MOID must be $<$ 0.00004263 au (one Earth's radius in au) and the 
     true anomaly associated with the orbit of the Earth must match the expected value at the collision time from the ephemerides. Assuming 
     Gaussian errors and in order to rank the computed candidate solutions, we use the following estimator:
     \begin{equation}
        \beta = {\Large e}^{-\frac{1}{2}\left[
                     \left(\frac{d - d^{*}}{\sigma_{d^{*}}}\right)^{2} +
                     \left(\frac{f - f^{*}}{\sigma_{f^{*}}}\right)^{2} 
                   \right]} \,,
                   \label{estimate}
     \end{equation} 
     where $d$ is the MOID of the test orbit in au, $d^{*}$ = 0 au is the minimum possible MOID, $\sigma_{d^{*}}$ is assumed to be the 
     radius of the Earth in au, $f$ is Earth's true anomaly used in the computation, $f^{*}$ is Earth's true anomaly at the collision time 
     and $\sigma_{f^{*}}$ is half the angle subtended by the Earth from the Sun (0$\fdg$00488). If $\beta > 0.368$, a collision is possible. 
     This technique gives the most probable orbit, statistically speaking, but degenerate solutions are possible, i.e. two or more most 
     probable orbits, but fairly different, can be found for a given collision event. This is consistent with the fact that, in theory, our 
     planet may suffer multiple, simultaneous impacts. In order to validate the method, we use data from the first and only case in which an 
     asteroid impact on the Earth has been accurately predicted, the Almahata Sitta event caused by the meteoroid 2008~TC$_{3}$ (Jenniskens 
     et al. 2009; Oszkiewicz et al. 2012). For this event, both the collision time (2008-Oct-07 02:46 UTC) and the pre-collision orbit ($a$ 
     = 1.308201 au, $e$ = 0.312065, $i$ = 2$\fdg$54220, $\Omega$ = 194$\fdg$101138 and $\omega$ = 234$\fdg$44897) are well 
     known\footnote{http://ssd.jpl.nasa.gov/sbdb.cgi?sstr=3430291} so a detailed comparison can be made. MOID results for 2008~TC$_{3}$ are 
     displayed on Fig. \ref{moid}, right-hand panels. The value of the MOID in au is colour coded following the associated colour box; the 
     best orbit appears as a green $\times$ sign. For this case, our algorithm gives a best orbit with $\beta = 0.987$ and parameters $a$ = 
     1.30807203 au, $e$ = 0.31187919, $i$ = 2$\fdg$5445889, $\Omega$ = 194$\fdg$091701 and $\omega$ = 234$\fdg$400944. If we average the 
     best 10 orbits ranked by $\beta$, we obtain $a$ = 1.3079$\pm$0.0006 au, $e$ = 0.3119$\pm$0.0004, $i$ = 2$\fdg$542$\pm$0$\fdg$006, 
     $\Omega$ = 194$\fdg$0918$\pm$0$\fdg$0011 and $\omega$ = 234$\fdg$44$\pm$0$\fdg$02 which agrees well with the orbit determined by Steven 
     R. Chesley above. Therefore, our primitive yet effective technique is robust enough to reproduce a well-established result that is 
     based on actual observations even if our method is purely geometrical. Using the same approach for the Chelyabinsk impactor, we obtain 
     a best orbit with $\beta = 0.9997$ and parameters $a$ = 1.62394517 au, $e$ = 0.53274620, $i$ = 3$\fdg$8210787, $\Omega$ = 
     326$\fdg$408990 and $\omega$ = 109$\fdg$466797 and the average of the best 10 orbits gives $a$ = 1.62375$\pm$0.00014 au, $e$ = 
     0.53279$\pm$0.00011, $i$ = 3$\fdg$817$\pm$0$\fdg$005, $\Omega$ = 326$\fdg$4090$\pm$0$\fdg$0007, $\omega$ = 109$\fdg$44$\pm$0$\fdg$03 
     and $M$ = 21$\fdg$99$\pm$0$\fdg$02, around the time of impact. This result matches well the orbit calculated by S. Nakano and included 
     in Table \ref{orbits}. Colour-coded MOIDs for the Chelyabinsk impactor are displayed in Fig. \ref{moid}, left-hand panels. The same 
     technique can be used to evaluate the statistical significance of the candidate solutions compiled in Table \ref{orbits}. If we 
     calculate the probability of obtaining a MOID under 0.05 au ($P_{\rm 0.050 au}$) and 0.00004263 au ($P_{\rm 0.00004263 au}$) at the 
     impact time (see above) and the highest $\beta$ rank for the various candidate orbits (see Table \ref{orbits}, last three columns), all 
     of them are statistically less robust than the one obtained here. In these calculations, the errors associated with the orbital elements 
     as provided by the respective authors (see Table \ref{orbits}) have been used when known; if unknown, the errors in Zuluaga et al. 
     (2013) have been used. From now on, we will assume that the orbit followed by the Chelyabinsk impactor prior to its collision was the 
     averaged one. 
     
%
%
%
%

  \section{Candidate selection}
     Now that the most probable orbit of the Chelyabinsk impactor has been established, the next step is finding candidates for the parent 
     body of the impactor as we assumed that it was the result of a relatively recent asteroid break-up event. Groups of rocky fragments 
     resulting from the disruption of an asteroid are called asteroid families. Trigo-Rodr\'{\i}guez et al. (2007) used the D-criterion of 
     Southworth \& Hawkins (1963), $D_{\rm SH}$, to investigate the connection between the asteroid 2002~NY$_{40}$ and the FN300806 bolide. 
     Alternatives to $D_{\rm SH}$ are the D-criterion of Lindblad \& Southworth (1971), $D_{\rm LS}$, which is based on the previous one and
     the $D_{\rm R}$ from Valsecchi, Jopek \& Froeschl\'e (1999). In order to investigate a possible association between the meteoroid 
     responsible for the Chelyabinsk superbolide as characterized by the orbital solution in Table \ref{orbits} and any known asteroid, we 
     carried out a search among all the objects currently catalogued by the JPL Small-Body Database\footnote{http://ssd.jpl.nasa.gov/sbdb.cgi} 
     using the three D-criteria. The lowest values of the various $D$s are found for 2011~EO$_{40}$ ($D_{\rm SH}$ = 0.12, $D_{\rm LS}$ = 
     0.011, $D_{\rm R}$ = 0.0084), followed by 2011~GP$_{28}$ (1.0, 0.015, 0.011),  2010~DU$_{1}$ (1.0, 0.020, 0.04), 2008~FH (1.0, 0.040, 
     0.049) and 2007~BD$_{7}$  (0.93, 0.047, 0.041) among those with relatively well known orbits. The best candidate, 2011~EO$_{40}$, was
     discovered on 2011 March 10 by Kowalski et al. (2011), has an absolute magnitude, $H$, of 21.7 (size of 150-330 m if $G$ = 0.15 and the 
     albedo range is 0.04--0.20) and it has been observed 20 times with an arc length of 34 d. The orbital elements of this Apollo asteroid 
     are $a$ = 1.6539$\pm$0.0004 au, $e$ = 0.54039$\pm$0.00014, $i$ = 3$\fdg$3638$\pm$0$\fdg$0008, $\Omega$ = 50$\fdg$310$\pm$0$\fdg$011 and 
     $\omega$ = 17$\fdg$055$\pm$0$\fdg$013. Apollo asteroid 2007~BD$_{7}$ was discovered on 2007 January 23, has $H$ = 21.1 and its orbit
     is based on 185 observations with a data-arc span of 14 d. Its orbital elements are $a$ = 1.5624$\pm$0.0012 au, $e$ = 0.4980$\pm$0.0005, 
     $i$ = 4$\fdg$849$\pm$0$\fdg$004, $\Omega$ = 343$\fdg$627$\pm$0$\fdg$008 and $\omega$ = 219$\fdg$875$\pm$0$\fdg$007. The other 
     candidates have shorter arcs and larger $H$: 2008~FH (12 d, $H$ = 24.4, $a$ = 1.582$\pm$0.012 au, $e$ = 0.504$\pm$0.005,
     $i$ = 3$\fdg$45$\pm$0$\fdg$03, $\Omega$ = 5$\fdg$207$\pm$0$\fdg$013 and $\omega$ = 264$\fdg$09$\pm$0$\fdg$04), 2010~DU$_{1}$ (4 d, $H$ 
     = 26.6, $a$ = 1.687$\pm$0.006 au, $e$ = 0.539$\pm$0.002, $i$ = 3$\fdg$704$\pm$0$\fdg$012, $\Omega$ = 147$\fdg$832$\pm$0$\fdg$002 and 
     $\omega$ = 74$\fdg$250$\pm$0$\fdg$010) or 2011~GP$_{28}$ (1 d, $H$ = 29.4, $a$ = 1.591$\pm$0.004 au, $e$ = 0.5199$\pm$0.0015, $i$ = 
     4$\fdg$048$\pm$0$\fdg$009, $\Omega$ = 16$\fdg$397$\pm$0$\fdg$003 and $\omega$ = 252$\fdg$19$\pm$0$\fdg$02). All these objects are 
     classified as Apollo asteroids, near-Earth objects (NEOs) and potentially hazardous asteroids (PHAs). Unfortunately, even the best 
     known orbits are based on rather short arcs. In the following, we perform $N$-body calculations to further study any possible dynamical 
     connection between some of the candidates and the superbolide. 
%
%
     \begin{figure*}
       \centering
        \includegraphics[width=\linewidth]{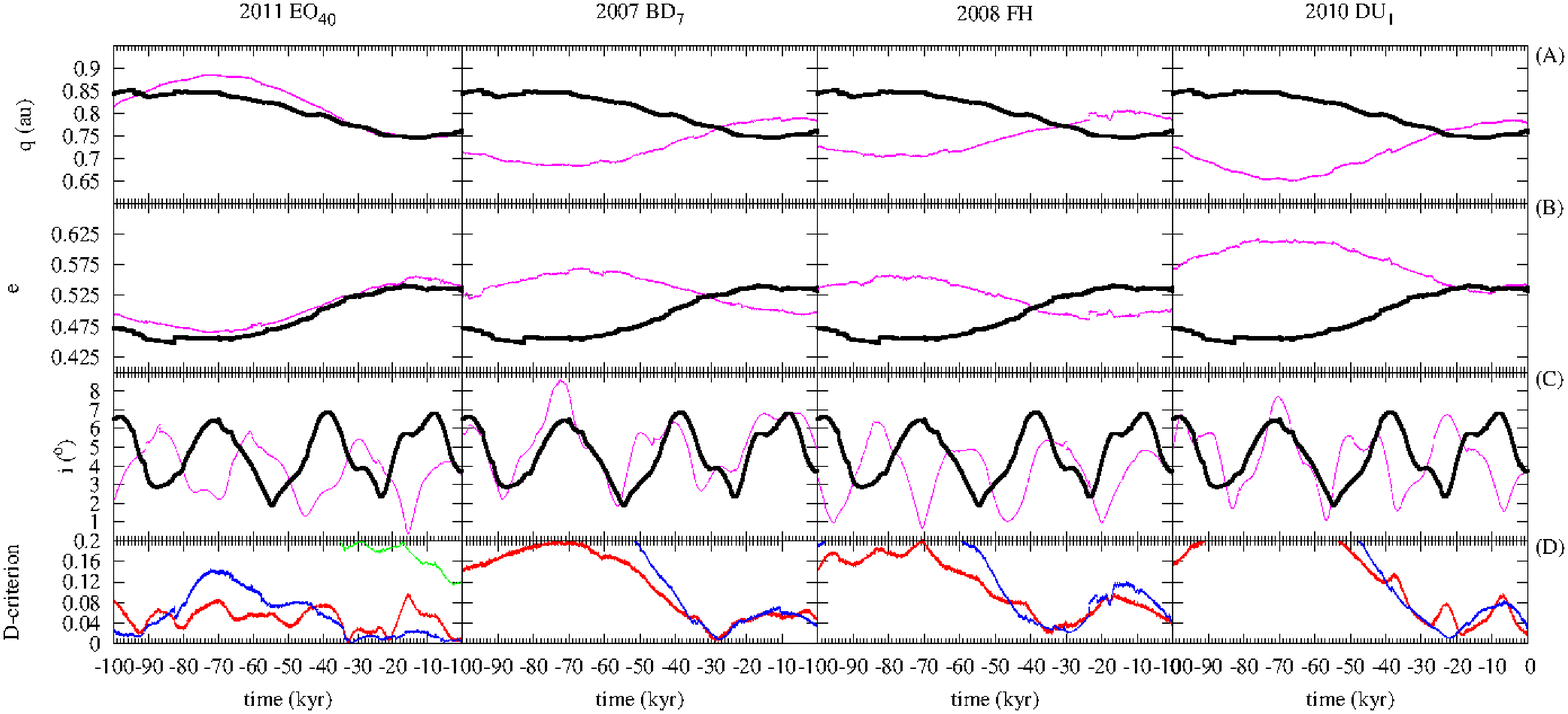}
        \caption{Results of the numerical integrations of the orbits backwards in time for 2011~EO$_{40}$, 2007~BD$_{7}$, 2008~FH and 
                 2010~DU$_{1}$ and one representative superbolide control orbit (see the text for details). The evolution of perihelia (A), 
                 eccentricities (B), inclinations (C) and the various D-criteria ($D_{\rm SH}$-green, $D_{\rm LS}$-blue, $D_{\rm R}$-red) 
                 are shown. The pink curve is always associated with the asteroid. The black curve shows the evolution of the control orbit 
                 of the meteoroid.
                }
        \label{f1}
     \end{figure*}
%
%

  \section{Dynamical evolution}
     The orbital evolution of meteoroid orbits following the osculating elements in Table \ref{orbits} (averaged orbit), those of the four 
     most promising candidate objects pointed out above and several others, were computed for 0.25 Myr backwards in time using the Hermite 
     integration scheme described by Makino (1991) and implemented by Aarseth (2003). The standard version of this serial code is publicly 
     available from the IoA web site.\footnote{http://www.ast.cam.ac.uk/$\sim$sverre/web/pages/nbody.htm} Results from this $N$-body code 
     have been discussed in de la Fuente Marcos \& de la Fuente Marcos (2012). Our direct integrations include the perturbations by the 
     eight major planets, the Moon, the barycentre of the Pluto-Charon system and the three largest asteroids. For accurate initial 
     positions and velocities, we used the heliocentric ecliptic Keplerian elements provided by the Jet Propulsion Laboratory online Solar 
     system data service\footnote{http://ssd.jpl.nasa.gov/?planet\_pos} (Giorgini et al. 1996) and based on the DE405 planetary orbital 
     ephemerides (Standish 1998) referred to the barycentre of the Solar system. In addition to the calculations completed using the nominal 
     orbital elements pointed out above, we have performed 50 control simulations for each object with sets of orbital elements obtained 
     from the nominal ones within the accepted uncertainties (3$\sigma$). Meteoroid orbits have been treated similarly. Fig. \ref{f1} 
     summarizes the results of our backwards integrations for the parent candidate asteroids. The orbital evolution of 2011~EO$_{40}$ 
     matches well that of the Chelyabinsk impactor. Giving the uncertainties in the initial conditions (orbital elements) for both 
     candidates and impactor, the agreement is good and suggests that these bodies were formed in a single (or a sequence of) break-up 
     event(s) 20-40 kyr ago. The orbits of these objects are strongly perturbed as they experience periodic close encounters not only with 
     the Earth--Moon system but also with Mars, Ceres and, in some cases, Venus. The objects studied here are part of a genetic family not a 
     dynamical one, like the NEO family recently identified by de la Fuente Marcos \& de la Fuente Marcos (2013). 

  \section{The Chelyabinsk asteroid family}
\begin{figure}
        \centering
        \includegraphics[width=\linewidth]{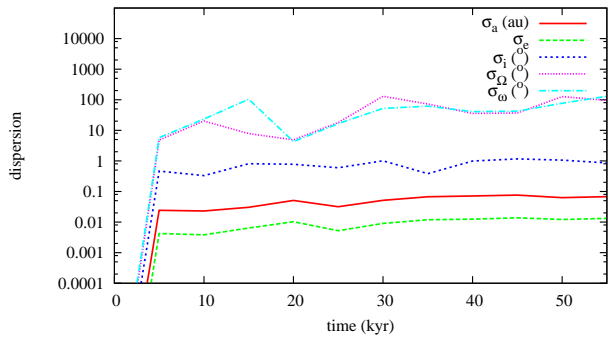}
        \caption{Time evolution of the dispersion of the orbital elements of a set of particles resulting from the smooth break-up of a 
                 rubble-pile asteroid as described in the text.
                }
        \label{cluster}
     \end{figure}

     So far, our numerical results are somewhat consistent with 2011~EO$_{40}$ and other minor bodies being members of a young asteroid 
     family but, can we identify additional members tracing a putative Chelyabinsk asteroid complex? and, what is more critical, can we 
     reasonably conclude that they could be the result of a break-up event? An analysis based on the various D-criteria shows about 20 
     candidates to be part of the proposed Chelyabinsk asteroid family. Unfortunately, most of them have $H >$ 25 and very short arcs (a few 
     days) so the actual characterization of the family is rather speculative although only objects with $D_{\rm R} < 0.05$ have been 
     tentatively selected. With this restriction, the orbital parameters of the proposed family and their spreads are $a$ = 1.66$\pm$0.08 
     au, $e$ = 0.54$\pm$0.02, $i$ = 3.7$\pm$1.3$^{\circ}$, $\Omega$ = 162$\pm$114$^{\circ}$ and $\omega$ = 173$\pm$96$^{\circ}$. In order to 
     check if these numbers are compatible with a gentle break-up event, we start a simulation with 100 test particles moving in orbits 
     similar to that of 2011~EO$_{40}$ but with negligible spread in their orbital elements at aphelion. This is equivalent to having a 
     smoothly disintegrating rubble-pile asteroid in which the relative velocities of the resulting fragments are basically zero. Fig. 
     \ref{cluster} shows the standard deviation of the various elements of the test particles as a function of the time. Although this 
     calculation gives no obvious constraint on the age of the family, the long-term values of their standard deviations match well the 
     values obtained above even if we do not consider non-gravitational effects that could be important for small objects. 
%
%
%
%

  \section{Discussion and conclusions}
     Although the numerical and statistical evidence in favour of a Chelyabinsk asteroid family or complex is quite encouraging, the 
     ultimate proof of a truly genetic relationship between all these objects requires spectroscopy or, much better, sample-return (e.g.
     Barucci et al. 2012). The analysis of the Chelyabinsk meteorites shows that they are chondrite breccias (Bischoff et al. 2013) so the 
     parent meteoroid may be of the S class. Our calculations did not include the Yarkovsky effect (see e.g. Bottke et al. 2006) which may 
     have a non-negligible role on the medium, long-term evolution of objects as small as the ones studied here. Proper modeling of the 
     Yarkovsky force requires knowledge on the physical properties of the objects involved (for example, rotation rate, albedo, bulk density, 
     surface conductivity, emissivity) which is not the case for the objects discussed here. Detailed observations during future encounters 
     with the Earth should be able to provide that information. On the short term, the Yarkovsky force mainly affects $a$ and $e$ but within 
     the dispersion range found here. Its effects are negligible if the objects are tumbling or in chaotic rotation. The non-inclusion of 
     this effect has no major impact on the assessment completed. 

     In this Letter, we have obtained a statistically robust solution for the pre-impact orbit of the Chelyabinsk superbolide. Assuming that
     such a meteoroid could be the result of the decay of an asteroid, we have singled out some candidates for membership in a putative 
     Chelyabinsk asteroid cluster or family and tested, using $N$-body simulations, their possible dynamical connection to the parent body 
     of the Chelyabinsk event. Our calculations suggest a dynamical link between some of the candidates and the superbolide but, 
     unfortunately, the current orbits of all the candidates are not reliable enough to claim a conclusive connection although the avaliable
     evidence is certainly encouraging. The situation is similar to that of PHA 2008~XM$_{1}$ recently studied by Madiedo et al. (2013).

  \section*{Acknowledgements}
     The authors would like to thank the referee, J.~M. Trigo-Rodr\'{\i}guez, for his constructive and helpful report and S. J. Aarseth for 
     providing a code used in this research. This work was partially supported by the Spanish `Comunidad de Madrid' under grant CAM 
     S2009/ESP-1496. We thank M. J. Fern\'andez-Figueroa, M. Rego Fern\'andez and the Department of Astrophysics of the Universidad 
     Complutense de Madrid (UCM) for providing computing facilities. Most of the calculations and part of the data analysis were completed 
     on the `Servidor Central de C\'alculo' of the UCM and we thank S. Cano Als\'ua for his help during this stage. In preparation of this 
     Letter, we made use of the NASA Astrophysics Data System, the ASTRO-PH e-print server and the MPC data server.

\end{document}